# Research Programs Arising from 'Oumuamua Considered as an Alien Craft

Martin Elvis[1]
Center for Astrophysics | Harvard & Smithsonian

## ABSTRACT

The controversial hypothesis that 'Oumuamua (1I/2017 U1) was an alien craft dominated by a solar sail is considered using known physics for the two possible cases: controlled and uncontrolled flight. The reliability engineering challenges for an artifact designed to operate for $\sim 10^5 - 10^6$ yr are also considerable. All three areas generate research programs going forward. The uncontrolled case could be either "anonymous METI" (messaging extraterrestrial intelligence) or "inadvertent METI". In the controlled case the nature of the origin star, trajectory guidance from the origin star to the Sun, and the identity of a destination star are all undecided. The "controlled" case has more strikes against it than the "uncontrolled" case, but neither suffers a knock-out blow, as yet. Some of the issues turn out not to be major obstacles to the alien craft hypothesis, but others weaken the case for it. Most, however, imply new studies. Some of these, e.g., intercept missions for new interstellar objects, are concepts being developed, and will be of value whatever these objects turn out to be. Overall, these considerations show that a many-pronged, targeted, research program can be built around the hypothesis that 'Oumuamua is an alien craft. The considerations presented here can also be applied to other interstellar visitors, as well as to general discussions of interstellar travel.

## 1. INTRODUCTION

'Oumuamua (1I/2017 U1) was the first unambiguously interstellar object to be discovered in the solar system (Meech et al. 2017). Controversially, it has been suggested to be an alien spacecraft (Bialy & Loeb 2018, Loeb 2021). Alien activity as an explanation for astronomical phenomena is often hard to follow up as these suggestions tend not to produce testable predictions. If researchers cannot see what research program an idea points to, then they have no incentive to change from their present research to this new field (Lakatos, 1980). This paper points out that a well-defined research program does arise for an alien craft interpretation of 'Oumuamua, leading to tests of this hypothesis that have various degrees of difficulty. This approach in no way rejects a natural explanation for 'Oumuamua, but simply aims at an impartial examination of the alien craft hypothesis by following the questions arising a few more steps. Though some estimates are made here, properly quantifying all the tests will require engaging a wide range of expertise.

The implications for the limited number of possible scenarios that flow from the assumption that 'Oumuamua is an alien craft are explored below. Katz (2021) has made some similar points, albeit in a less open-ended fashion. The same general considerations will apply to any future objects for which this hypothesis is raised.

The argument in favor of the alien craft hypothesis can be readily summarized. 'Oumuamua clearly accelerated on its way out of the solar system, a measurement made at high confidence (Micheli et al., 2018, but see Rafikov 2018 and Katz 2019). The acceleration was small, $\sim 5 \times 10^{-6}$ m s$^{-2}$, but highly statistically significant (30$\sigma$, Micheli et al., 2018.) The acceleration was not constant but dropped off with heliocentric distance close to a $1/r^2$ law. No cometary dust (Meech et al., 2017) or gas (for CO and $CO_2$, Trilling et al., 2018) ejection was observed that could cause the observed non-gravitational radial acceleration. If no mass was lost, then the remaining possibility is that 'Oumuamua was accelerated by solar radiation pressure and the solar wind. This is prima facie plausible as the large (1.5 mag, a factor ~4) amplitude of the 'Oumuamua light curve implies a highly elongated or highly flattened geometry, a cigar or a pancake with a large, but not unprecedented, 6:1 axial ratio (Meech et al, 2017, Trilling et al. 2018, Mashchenko 2019). A large area for intercepting solar radiation may then be available. This alone does not imply a non-natural explanation. However, to achieve the observed acceleration using these weak forces 'Oumuamua would have to have a huge area/mass ratio ~0.1 g $cm^{-2}$, due to either low density, ~$10^{-5}$ g $cm^{-3}$, or extreme thinness, ~0.3 – 0.9 mm, giving a resultant mass of only ~750 kg for this ~100 m scale object (Bialy & Loeb 2018.) An extremely thin shape is suggestive of an artifact to some researchers.

---

[1] ORCID ID: 0000-0001-5060-1398

Natural explanations are not at all ruled out. Several astrophysical suggestions have been made to explain the unusual shape, high area/mass ratio, and acceleration of 'Oumuamua. E.g., a fractal icy aggregate with a density of ~$10^{-5}$ g $cm^{-3}$ (Moro-Martin 2019, Luu et al. 2020); a comet fragment (Sekanina 2019); molecular hydrogen ice (Seligman & Laughlin 2020); molecular nitrogen ice (Jackson and Desch 2021); a fragment from a planetoid destroyed in a tidal disruption event around a white dwarf (Rafikov 2018); and some have been critiqued (Levine et al., 2021, Hoang & Loeb 2020, Loeb 2021, Siraj & Loeb, 2021, Phan, Hoang & Loeb 2021).

While most scholars find the astrophysical options plausible, for some researchers the evidence against the proposed astrophysical explanations is compelling. For these researchers the only remaining possibility is that 'Oumuamua is dominated by a thin, large area, "solar sail" constructed by an alien civilization (Loeb 2021). This proposal has itself been highly criticized (Katz 2021).

The approach taken here is to investigate the consequences of an alien spacecraft explanation so that these consequences can be studied further to either reinforce or reduce the likelihood of this option. The number of possibilities for this option appears to be quite limited, allowing the analytic approach used here.

Unknown technologies or physics for the alien craft are not invoked here, as that approach is unconstrained. Instead, the assumption is that the putative alien designers worked with the same physics we have, the same fundamental constraints on materials that we have, and design constraints that are familiar to us. The result is that, for any of the cases explored, their craft must perform to some demanding and sometimes well-constrained capabilities.

In the alien spacecraft hypothesis either 'Oumuamua is piloted (not necessarily by alien life directly but by their computers) and so is under control, or it is derelict, i.e., not under control. The long operational lifetime of 'Oumuamua in this hypothesis raises further issues. All three areas have consequences that are explored here.

## 2. UN-GUIDED OR DERELICT

The simpler of the two possibilities is that 'Oumuamua is un-guided. 'Oumuamua was found to have a rotation period of ~7 hours, though the light curve did not repeat precisely indicating a tumbling motion (i.e., non-principal axis rotation, Meech et al. 2017, Drahus et al. 2018, Fraser et al. 2018). This rotation state suggests that it is not under control, or not wholly under control. The object was then behaving like a derelict vessel. 'Oumuamua entered the solar system with a velocity very close to that of the Local Standard of Rest (LSR, Mamajek, 2017.) As 'Oumuamua is in the LSR it is more correct to say that the Sun travels at this speed in the LSR. Any option requiring 'Oumuamua to be under control is weakened, given this drifting trajectory.

## 2.1 Anonymous METI

'Oumuamua may not be truly derelict but may instead be a passive "message in a bottle" (Loeb 2018). In this hypothesis 'Oumuamua was designed to let technological life (because telescopes were needed to detect 'Oumuamua) around another star realize the existence (at least at one time) of another civilization. The solar sail, in this case, may have been intended not for realistic propulsion but to give a small but tell-tale acceleration. That would require that the sail orient itself at times perpendicular to the radial direction from the Sun. A tumbling motion will reduce the amount of time the sail is oriented correctly to achieve a significant acceleration. If the sail were always oriented radially then 'Oumuamua would have decelerated on approaching the Sun. That possibility would affect the apparent incoming direction which could modestly affect the arguments in Sec.3.1 below.

An additional possible motivation to situate a spacecraft in the LSR would be to deliberately hide its origin. The message in the bottle would then be "You are not alone, but we are not telling you where we are." Such a design may have been motivated as being a safe version of messaging extraterrestrial intelligence ("Anonymous METI"[2]) In this case we would expect that a mission to 'Oumuamua, or another suspiciously behaving interstellar object (e.g. Hibberd & Hein, 2021), would find clear evidence of these bodies being a deliberate construct, but would have no identifying marks equivalent to the plaques mapping out Earth's location with respect to nearby pulsars on Pioneers 10 and 11 and Voyagers 1 and 2[3].

## 2.2 Inadvertent METI

The upper stages of the launchers that put our space probes onto their final trajectories also reach a similar trajectory. There are numerous upper stages in heliocentric orbits and a small number on solar system escape trajectories, e.g., the New Horizons upper stage (J.C. McDowell, 2021, private communication.) These are inadvertent interstellar spacecraft, and so a form of unintended METI from the sender's perspective. From our perspective they are another example of a technosignature.

Similarly, the putative builders of 'Oumuamua may have manufactured these craft as upper stages, presumably accelerated by directed energy (e.g., lasers), given their light-sail like design. This form of acceleration is a good way to overcome the limits imposed by the rocket equation. In this case there must have been a highly active stellar system economy around the origin star $\sim 10^5$ yr ago. Possibly this activity would leave other tracers, for example from the mining of asteroids for the raw material from which to build the craft (Forgan & Elvis, 2011.) These "craft" may even be just tailings from mining in the stellar system's Oort cloud. These tailings would have been given a small velocity to expel them into interstellar space, perhaps in order to reduce threats to their infrastructure in their Oort cloud. Very approximately, comets at $10^4$ AU will be moving at ~30 m s-1, sufficient to damage loosely bound objects (including spacecraft). The space density of Oort cloud comets from $10^4 - 10^5$ AU is only ~1 per 4000 AU$^3$ making collisions rare, but if say 10% of the comets were cut into ~1 t slices there would be ~$10^{21}$ of them (Weissman 1983) for a space density of ~$10^5$ AU$^{-3}$ a separation of ~$5\times10^6$ km, that they would traverse in ~5 years. Though their cross-section is ~0.01 km$^2$, and their probability of collision ~$2\times10^{-15}$ there would still be ~$10^7$ collisions/year. For a civilization with large amounts of infrastructure in their Oort cloud this could be sufficient motivation to eject the comet slices. A true runaway "Kessler Syndrome" (Kessler & Cour-Palais, 1978) may not develop.

---

[2] anonymous messaging of extraterrestrial intelligence.
[3] JPL: https://voyager.jpl.nasa.gov/golden-record/golden-record-cover/ (accessed 5 March 5, 2021)

The "message" from inadvertent METI would be simply "we exist" with any other information gleaned from a close inspection being also inadvertent.

## 2.3 Population Implications

As 'Oumuamua has a ~$10^5 - 10^6$ yr travel time (see Sec.3.1), it was constructed before any detectable human technology was created, possibly even before homo sapiens evolved. This implies that the "message in a bottle" strategy must have been implemented without knowing which stars would have technological societies capable of detecting 'Oumuamua. There must then be at least a moderately large number of copies in order to swing through the nearby solar systems at a rate that makes their detection likely given a distribution of onset times and lifetimes for this capability. The LSR velocity argues against efficient targeting, keeping the total number large.

If 'Oumuamua came into the Solar System entirely by chance, without even the "message in a bottle" plan behind it, then the population density of such objects, at least locally, would be the same as if it were a naturally occurring object, and interstellar asteroid. On this basis Do, Tucker and Tonry (2018) estimate the space density of 'Oumuamua-like objects to be 0.2 AU$^{-3}$ or 2 x 10$^{15}$ pc$^{-3}$. Loeb (2018) says that "This would require the unreasonable rate of a launch every five minutes from a planetary system even if all civilizations live as long as the full lifetime of the Milky Way galaxy." This high rate has been taken as arguing for a natural origin.

The maximum total mass of all copies of 'Oumuamua in this scenario would be ~$10^{14}$ t (metric tonnes, for a density ~$10^{-4}$ g cm$^{-3}$). While far larger than the Earth's proven iron reserves[4] (2.3x10$^{11}$ t), this mass is small compared with, e.g., the ~$10^{18}$ t of iron in the Main Belt asteroids (Elvis and Milligan, 2019), and deployment could well have taken place over 1% of the planned $10^5$ yr lifetime, or ~$10^3$ yr, as this implies a launch rate of ~3000 s$^{-1}$. For an Earth-scale economy this is an unattainable number, but for a solar system scale economy, where the Main Belt asteroids are fully exploitable, many thousands of 'Oumuamua factories and launchers are within the range of plausibility.

The putative designers may well have known which planets harbored, or could harbor, complex life. As roughly a quarter of all M-dwarfs have rocky planets in their habitable zones (Dressing & Charbonneau, 2015), this knowledge may not reduce the number of alien craft required by a large factor. As we learn more about the habitability of exoplanets the fraction may go down. For example, stellar activity may remove the atmospheres of all Earth-sized planets in the habitable zones of M stars (Van Eylen et al., 2018) The total numbers and the required launch rate may then be reduced by some orders of magnitude if the putative builders had this knowledge.

---

[4] https://minerals.usgs.gov/minerals/pubs/commodity/iron_ore/mcs-2017-feore.pdf (accessed 22 March, 2021.)

Ongoing and expanded surveys (e.g., VRO/LSST[5], Moro-Martín et al., 2009) for more interstellar asteroids will determine population density of interstellar objects of whatever kind and so put constraints on this case.

## 3. UNDER CONTROL

If 'Oumuamua was under control during the solar system transit, then all the properties of 'Oumuamua are deliberate and require investigation. The first question is why is it in the LSR if the Sun is its destination? Additional questions immediately arise about (1) the origin and (2) the destination of 'Oumuamua, and (3) the achievable accuracy of the inbound trajectory. As the design is dominated, in this hypothesis, by a solar sail, the craft was surely using that solar sail to navigate when near a radiation source.

**3.1 Where did it come from?** 'Oumuamua must be coming from some other star if it uses the light pressure from that star to accelerate. As the light sail dominates the system mass then to be justify that mass, the origin star ought to be nearby, so that 'Oumuamua does not spend most of its time in the isotropic illumination of interstellar space, where the solar sail is of no use. Active reflectivity control of the sail could continue to generate a modest acceleration. Otherwise, the use of a light sail would be a poor engineering choice. In addition, the challenge of lifetimes $>10^6$ yr are significant, especially for large thin structures (see section 4) arguing for a local origin. There are several possibilities for the origin star.

'Oumuamua had a hyperbolic orbit[6] (e = 1.20), arriving from (RA, dec) = (18h 42m, +34.3º, ± 5'), with an asymptotic velocity of 26 km s$^{-1}$ (Meech et al., 2017, Bayler-Jones et al., 2018). At this speed 'Oumuamua travels 2.5 pc in $10^5$ years.

### 3.1.1 Main Sequence Stars

No cataloged stars lying within 60 pc of the Sun come within one parsec of the trajectory, and only four stars come within 4 pc in SIMBAD (Dybczyński & Królikowska, 2018), and only 6 stars come within 2 pc in the deeper Gaia DR2 catalog over the past $10^6$ yr (Bayler-Jones et al., 2018). These papers assume no maneuvers before the detection of 'Oumuamua. The minimum course corrections needed to create a rendezvous with another star in the past would be useful to calculate.

Only stars with main sequence lifetimes of ~1 Gyr or more are likely to have developed complex life. By terrestrial standards 1 Gyr is barely enough time to harbor multi-cellular life. The oldest certain fossils found so far on Earth come from a time when the Earth was ~1 Gyr old (Schopf et al. 2017). There are suggestions of single cell life as early as 0.3 Gyr (Dodd et al., 2017.) Complex life, however, seems to have taken much longer, ~4 Gyr, although the Cambrian explosion that happened then was rapid (Parry et al., 2017). Hence, stars with masses $>2.5$ $M_{sol}$ that have Main Sequence lifetimes < 1 Gyr, are unlikely origin stars (spectral type A2 or earlier). This concern, though, assumes that Earth is a typical example, which it may not be as we do not know what triggered the Cambrian explosion.

### 3.1.2 Brown Dwarfs

---

[5] Vera C. Rubin Observatory: https://www.lsst.org/scientists
[6] JPL Small Body Database: https://ssd.jpl.nasa.gov/sbdb.cgi#top (accessed 26 March 2021.)

Could 'Oumuamua have come from an undetected star within ~10 light-years? The obvious, and quite numerous, population of quite hard to detect stars would be brown dwarfs, i.e. T and L stars. T and L stars are long-lived enough to potentially develop life in their systems.

These stars have been searched for in infrared surveys with UKIDSS and WISE, and in deep optical imaging for high parallax and proper motion (Best et al., 2021). None presently lie within 1 degree of the origin point of 'Oumuamua.

Brown dwarfs have surface temperatures, $T \sim 1000$ K, versus 6000 K for the Sun. As the emitted black body power goes as $\sigma T^4$, a brown dwarf will produce just $\sim 10^{-3}$ of the acceleration that the Sun provides. The launching process would then have to be by other means. An electric sail in one option (Lingam & Loeb, 2019). A laser that reflects off the sail would be an efficient design choice, as for the Breakthrough Starshot[7] project. The energy required to accelerate the mass of 'Oumuamua (~1 t) to ~20 km s$^{-1}$ is $2 \times 10^{11}$ J, so a 1 GW laser would have to operate for 200 s to achieve this effect. This is not far from the Breakthrough Starshot power and duration requirements.

### 3.1.3 Origin Star went Supernova

Could the star of origin have gone supernova since 'Oumuamua departed ~$10^5$ years ago and so no longer be present?

Supernova remnants are detectable up to ~$10^5$ years after the event (e.g., Cygnus Loop at $2 \times 10^4$ yr, and a radius of 18.5 pc, Fesen et al., 2018) and so may still be visible, depending on how soon after the departure of 'Oumuamua the supernova happened. If a neutron star was created in the supernova, then that object should still be visible as a thermal source at ~$10^6$ K (Bignami 1996), though it may have moved significantly in that time (~250 pc for a typical ejection speed of 250 km s$^{-1}$, Faucher-Giguère & Kaspi, 2006), in an unknown direction. Old supernova ejecta shells travel an order of magnitude faster than 'Oumuamua (~200 km s$^{-1}$, Fesen et al. 2018) and so would have traveled well past the Sun and would now lie in the opposite direction to the 'Oumuamua origin point and would have a large angular size.

Detection of large diameter supernova remnants via radio and UV emission has been improving. Fesen et al. (2021) report three strong candidates at distances of ~100 pc. Finding convincing evidence of a ~10 pc distant supernova remnant will be challenging but could be attempted. Abundance measurements of the surfaces of asteroids might show recent enhancements of $^{26}$Al and other short-lived radioactive nuclei (with half-lives of $10^5 - 10^7$ yr) characteristic of Type II supernova (Adams, 2010) The samples returned by Hayabusa 2 and OSIRIS-REx should be able to test this scenario. Massive stars tend to form in groups and there are only two plausible groups within 25 pc, B Pic and AB Dor (Mamajek, 2015); neither is close to the origin direction of 'Oumuamua.

A high mass star is unlikely to have been the location where the craft was built. The massive ($8 < M < 40$ M$_{sol}$) stars that create core-collapse type II supernovae are short-lived (<3 x $10^7$ yr, Woosley et al. 2002). The Earth was mostly molten for the first ~$10^7$ yr after the Moon-creating impact, the Hadean Eon (Sleep, Zahnle and Lupu 2014). It is hard to imagine life, let alone a technological civilization, growing under such conditions and so is highly unlikely within the lifetime of the star. In that case the supernova star would have been only a way station for 'Oumuamua.

---

[7] https://breakthroughinitiatives.org/initiative/3 (accessed 26 March 2021.)

This option raises other questions beyond the scope of this paper: how much would the supernova light and ejecta accelerate 'Oumuamua given the sail area (see Sec. 4 of Do et al., 2018)? How far from the supernova would 'Oumuamua have to be for the sail to survive the ejecta passage at velocities of 1000 km s$^{-1}$ or more.

### 3.1.4. Compact Objects as Way Stations

'Oumuamua may not have come to us not from the star system where it was built, but instead after making a slingshot gravity assist maneuver around another object that was merely a way station for the craft. Why a way station would be used to deliver 'Oumuamua into the LSR rather than a swifter trajectory is a puzzle.

Compact objects, either an isolated neutron star or a black hole, are particularly effective gravity assist tools in order to set course for the Sun.

A neutron star is a plausible candidate as the expected space density of neutron stars local to the Sun is predicted to be ~3.6 x 10$^{-4}$ N$_9$ pc$^{-3}$. N$_9$ is the total number of isolated neutron stars in the Galactic disk in units of 10$^9$, that is the current best estimate (Sartore et al, 2011.) This implies that the most likely distance to the nearest neutron star is ~10 pc (Sartore et al, 2011). With some luck then an isolated neutron star could have been well-positioned to serve as a gravity assist body for 'Oumuamua. Isolated neutron stars often acquire a high velocity from the supernova event in which they formed, ~250 km s$^{-1}$ (Sartore et al., 2011)  This makes them attractive as gravity assist targets (Johnson 2003), although the LSR-like velocity of 'Oumuamua suggests only a weak velocity boost.

Isolated black holes are estimated to be about an order of magnitude rarer than isolated neutron stars with ~10$^8$ in the Galactic disk and bulge (Fender et al., 2013.) For a uniform disk distribution (a more naïve calculation than that of Sartore et al. 2011), Fender et al. (2013) find a density of 10$^{-4}$ pc$^{-3}$, and a mean separation of just over 10 pc, so again it is plausible that an isolated black hole was conveniently located for 'Oumuamua.

For both types of compact object, the hypothetical trajectory designers need to have mapped out the local population of both types of object in position and velocity to high enough precision to ensure that the encounter was at the right impact parameter to produce an outbound trajectory that would give an impact parameter with the Sun of ~0.25 AU, as discussed in Sec. 3.2 below.

Both neutron stars and black holes are hard to detect when isolated. Limits can be put on both types of object in soft X-rays and in the EUV. Within ~100 light-years interstellar absorption is low, and a handful of unidentified EUV candidates exist (Maoz et al., 1997.) These may be neutron stars, but none lies near the 'Oumuamua origin point. At a typical space velocity of 250 km s$^{-1}$ (Faucher-Giguère & Kaspi, 2006), the putative neutron star of origin for 'Oumuamua will have traveled 25 pc in 10$^5$ yr, or up to 2.5° on the sky at a distance of 10 pc, depending on the direction of travel. Searches using the eROSITA all-sky survey instrument on Spektr-X-gamma (Predehl 2017), and with more targeted soft X-ray and EUV instruments can be undertaken.

Bialy & Loeb (2018) estimate the tensile stress on a body due to tidal forces, P$_{tid}$, will be ~10$^{-6}$ r$_{AU}$ d$_4^2$ M/M$_{sol}$ r$_{AU}^{-3}$ dyne cm$^{-2}$, where r$_{AU}$ is the distance of 'Oumuamua from the star (the Sun in the case considered by Bialy & Loeb) in AU, d$_4$ is the length of 'Oumuamua in units of 10$^4$cm, and M/M$_{sol}$ is the stellar mass in solar units. At an altitude of 300 km (~2x10$^{-6}$ AU) above an M/M$_{sol}$ = 1 compact object, P$_{tid}$ ~ 1x10$^{11}$ dyne cm$^{-2}$. This is comparable to the tensile strengths of diamond or monocrystalline silicon (see Table 1 in Bialy & Loeb 2018). These theoretical limits to material strengths imply that there is a closest possible approach to both a neutron star and a black hole, which will limit the gravity assist maneuvers achievable.

## 3.2 Course navigation

How accurately can a spacecraft be navigated using only an initial external radiation powered phase? Is that accuracy sufficient to make a precision "landfall" within a fraction of an AU at a target star ~10 parsecs away? An error of 1 arcsec at 10 pc is 1.5x10$^9$ km (10 AU), compared with the perihelion of 'Oumuamua of 3.7x10$^7$ km (0.255 AU[1]). To reliably put 'Oumuamua on a solar gravity assist course the trajectory must then be accurate to the milli-arcsecond level. If instead the goal was only to get close to Earth, as a complex life-bearing planet, but avoid the Sun, then 0.01 arcsecond, or even 0.1 arcsecond, accuracy may suffice.

Solar sails employing starlight are only effective out to ~130 AU from a solar type G star, after which the solar the radiation field becomes <1% of the diffuse Galactic optical background light (~10$^{-9}$ erg s$^{-1}$ cm$^{-2}$ sr$^{-1}$ Å$^{-1}$, 0.2 – 1 μm, Bernstein, Freedman & Madore, 2002) and the effective acceleration becomes negligible. Later course corrections are not possible until the radiation from the target star becomes significant, again at ~130 AU for a G star. A similar lack of course correction arises for the gravity assist case. Alternatively, the sail may be powered by a laser from farther back along the trajectory to enable a course correction. The feasibility of a laser system designed to work parsec over distances are demanding and should be investigated.

Achieving milliarcsecond accuracy in a gravity assist is particularly challenging, as the deflections gained from gravity assists are highly sensitive to the precise impact parameter of the incoming object from the massive object (e.g., Yefremov 2020.) The designers would have needed to know the 3D locations and space velocities of the compact object target to high accuracy in order to achieve the correct impact parameter ~10$^5$ yr later. It would be interesting to determine whether there are fundamental limits to the achievable accuracy, given even small perturbations from other objects along the path, e.g., comets, interstellar asteroids.

Course corrections could be made at a later point in the voyage, when the needed corrections could be more accurately measured. Such corrections would be correspondingly larger and would need a supplemental power source. A 1 arcsecond trajectory error can be corrected with a delta-v = 13 cm s$^{-1}$ for a 26 km s$^{-1}$ velocity. For 'Oumuamua, with a mass of ~1 mt (Bialy & Loeb 2018), this would require just 16.9 J. A one degree correction of a 1 mt payload would require 6.1x10$^4$ J for a delta-v of 0.49 km s$^{-1}$, or 100 W for 5 minutes. This is not a demanding requirement. For a typical chemical rocket exhaust velocity of 10$^4$ m s$^{-1}$, the rocket equation gives a fuel mass of 120 kg. A rough limit to the correction would be reached when the fuel mass is equal to the payload mass, which happens when delta-v = 6.9 km s$^{-1}$. Higher exhaust velocities, e.g., from ion engines, would allow larger delta-v within the mass limit. A zero propellant, albeit slow, way to change course would be for the light sail to change its albedo to be high on one side and low on the other, leading to an asymmetric force even in a uniform background (e.g., Ehresmann 2021.)

## 3.3 Where is it going?

'Oumuamua is departing towards (RA, dec) = (23h 51m 24.3s, +24º 42' 59.0") (Bayler-Jones et al., 2018). This exit asymptote from the solar system does not point toward any known nearby star[8] (Best et al., 2021).

'Oumuamua will have a distant (~0.46 pc) encounter at ~100 km s$^{-1}$ with the currently 3.1 pc distant M5 dwarf Ross 248 in 29,000 years (Bayler-Jones et al., 2018). A ~10 degree course correction early in the flight toward Ross 248 would bring 'Oumuamua to a close encounter with Ross 248. The acceleration of 'Oumuamua around the Sun (Micheli et al., 2018) altered the outbound direction by of order a few arcminutes compared with an extrapolation of the inbound hyperbola (Bayler-Jones et al., 2018.) A degree sized maneuver was then plausibly within the capability of 'Oumuamua, if it is an alien craft. Unfortunately, we do not now have the means to observe such a course change. Perhaps the solar encounter was a failure? This would contradict the extreme accuracies invoked earlier.

As in the case of the incoming trajectory, 'Oumuamua could be heading towards an undetected compact object for a gravity assist. The space density of these objects is low enough that invoking one at both ends of the trajectory seems extreme. However, they can be searched for in the same way as for the incoming trajectory.

Without a destination the large deflection of 'Oumuamua as it rounded the Sun becomes another unexplained, low probability, event, and so argues against the under-control alien craft hypothesis.

## 4. RELIABILITY ENGINEERING CONSIDERATIONS

### 4.1 Lifetime

The reliability engineering challenges (see e.g., O'Connor and Kleyner, 2012) in designing to a $10^5$ yr lifetime are daunting. This lifetime is ~20x longer than the oldest human-built structures on Earth, most of which, though massive and built from robust materials, have suffered significant damage, albeit in a challenging environment. More relevant is that this lifetime is ~1000x longer than the lifetime of any industrial revolution machine, and ~4000x longer than the longest-lived spacecraft, Voyagers 1 and 2 launched in 1977.[9]

At 10,000 years, the "Clock of the Long Now" has the longest design lifetime of any current technology[10]. The lifetime of the Clock of the Long Now is a demonstration that plausible design lifetimes ~100 times greater than existing machinery can be conceived of. The lifetime needed for 'Oumuamua if it is an alien craft is 10x longer, which is however a relatively small factor. Studying the Clock of the Long Now principles may show ways that lifetimes comparable with the 'Oumuamua journey time could be realized. If 'Oumuamua came to us from a way point, rather than from the site of its construction, then all the lifetime requirements need to be at least doubled.

---

[8] SIMBAD search "region (23.7 +24.8, 1d) && plx > 40" (accessed 22 March, 2021.)
[9] NASA JPL: https://voyager.jpl.nasa.gov/mission/ (accessed 22 March 2021.)
[10] The Long Now Foundation: https://longnow.org/clock/ (accessed 1 April 2021.)

The challenge of reliable operation over ~$10^5$ yr means that the mean time to failure (MTTF), a measure often used to analyze failures in complex systems, is ~5x longer, 500,000 years. This timescale arises from the result that for a non-repairable system to have 99% probability of surviving to a given time (assuming Poisson failure distributions), engineers need to design for a MTTF five times longer[11]). Hence the designers would need to have designed for a MTTF approaching $10^6$ yr for 'Oumuamua, ~100x longer yet than for the Clock of the Long Now. Studies made for the 100 Year Starship may be helpful starting points[12]. Failing to find solutions would argue against the alien craft hypothesis

### 4.1.1 Materials stability

The designers of 'Oumuamua would be constrained to the same elements that we have, although they would surely have compounds novel to us. It would be interesting to see whether we could conceive of designs that could approach these lifetimes given theoretical limits on materials properties especially for large thin structures such as light sails. For example, Bialy & Loeb (2018) show that a light sail made of diamond or silicate material, with tensile strengths of ~$10^{10}$ dyne cm$^{-2}$ could lose about half of its mass due to interstellar dust bombardment. That is quite a narrow engineering margin, so the sail may well fail before any post-solar stellar encounter. If the craft used a gravity assist around another stellar body, then the time spent by the craft in interstellar space will have been of order a factor two longer, endangering the sail. In addition, the environment around the stellar gravity assist body may well have a higher density of dust. Similarly, encounters with zodiacal dust in the solar system would be relatively short-lived, but at higher speed and particle density. More sophisticated modeling using detailed material properties would be valuable.

Diamond sheets of this type can be made but large single crystal sheets are well beyond our technology today (Butler & Sumant 2008), though not theoretically ruled out. However, graphite is a lower energy state than diamond and diamond will convert to graphite at raised temperatures or under ion bombardment (Setton, Bernier, and Lefrant 2002). Dust impacts will raise temperatures locally (on a ~mm scale), while cosmic ray bombardment is continuous. Estimates of the lifetime of diamond sheets under the conditions of interstellar space would be instructive. In principle superconducting magnets could be used to create an artificial magnetosphere[13] to shield the light sail from cosmic rays. To keep within the mass budget these magnets must be low density and/or thin.

Silicon monocrystals are also subject to degradation at elevated temperatures and irradiation (Sperber 2019), but investigations so far have been in the context of photovoltaic cells. Investigation of monocrystalline silicon under interstellar conditions would also be of interest.

### 4.2 Repair

For repairable systems the appropriate metric is the less demanding mean time between failures (MTBF, O'Connor and Kleyner, 2012). Repair raises its own issues. No outside materials would be available for 'Oumuamua, so the craft would require recycling. Recycling has limits; material ablated off the thin sail by collisions with interstellar dust, which could be half of the initial mass (Bialy & Loeb 2018), would be lost. Repair would require a supply of raw material on board, and complex mechanisms to carry out the repairs, reducing available payload mass; the repair mechanisms themselves must have sufficiently long MTBFs. These mechanisms may have to operate after ~$10^5$ years of inaction, which presents other challenges. Repair also implies a power source during interstellar cruise, though the amount of power may be quite small. A fault detection mechanism would seem to imply sensors and a computer that can operate for $10^5$ yr. As cosmic ray hits accumulate with time the computer must have multiple redundancies for error avoidance and likely thick connectors to survive. The necessary "super rad hard" standards could be estimated.

---

[11] Arenberg, J., 2021, slide 21, Ivy Space Coalition Conference, February 6-7, private communication.
[12] 100 Year Starship: http://www.100yss.org (accessed 5 March 2021.)
[13] Rutherford Appleton Lab. 2015: http://www.minimagnetosphere.rl.ac.uk

Katz (2021) notes that tumbling motions will be damped out by flexing in the sail structure, and so the observed tumbling would not be expected. The sail may not flex, though, as it could have been constructed on-orbit where no deployment mechanism is required, and could then be a single (e.g., 3D printed) structure.

Keeping key elements of the craft sufficiently warm with radioisotope heating would require the use of radioactive materials with half-lives comparable to the journey time, ~$10^5$ yr, greatly restricting the number of radionuclides available[14] and requiring larger mass to obtain the same power that more rapidly decaying materials would provide. Are there alternative low-mass long-lived heat sources? Otherwise, the temperature of the craft could be allowed to be very low. Low temperature operation may have benefits from, e.g., superconducting components, though mechanisms operating at a few K may face fundamental limits.

## 5. DISCUSSION AND CONCLUSIONS

The above discussion highlights a number of challenges for the alien craft hypothesis that further analysis could examine. Table 1 summarize the cases, the issues they raise, and the implications and/or tests that flow from these issues. Figure 1 puts these in the context of the broader picture including astrophysical explanations.

The "controlled" case has more strikes against it than the "uncontrolled" case, but neither suffers a knock-out blow, as yet. Some of the issues turn out not to be major obstacles to the alien craft hypothesis, but others cast doubt on it. All of the discussion here is highly simplified. Most of the issues suggest new, more sophisticated, studies that could be carried out. Some of these, e.g., intercept missions for newly discovered interstellar objects, are concepts already being developed and will be of value whatever these objects turn out to be.

Overall, these considerations show that a broad and well-defined research program can be built around the hypothesis that 'Oumuamua is an alien craft. More generally, the considerations presented here can also be applied to other interstellar visitors, as well as to general discussions of interstellar travel.

The author acknowledges helpful contributions from J. Arenberg, C.J.F. Elvis, G. Fabbiano, A. Lawrence, J.C. McDowell, M. Micheli, J. Wright, and an anonymous referee.

---

[14] CERN: A Table of Frequently Used Radioisotopes: https://cds.cern.ch/record/1309915/files/978-3-642-02586-0_BookBackMatter.pdf (accessed 1 April 2021.)

Table 1: Summary of Cases, Issues, and Tests (section numbers are in parentheses.)

| Case | Issue | Implication/Test |
|---|---|---|
| Uncontrolled (2) | | |
| Anonymous METI (2.1) | Tell-tale acceleration | Tumbling implies lower time for useful orientation. Needs estimating. |
| | No identifying marks | Close-up inspection. Presently not possible for 'Oumuamua; possible for future interstellar objects. Missions valuable whatever their nature. |
| Inadvertant METI (2.2) | Signs of large-scale mining $10^5$ yr ago, or ongoing, around origin star | Close-up inspection. Identifying marks. Unexpectedly large debris disks in older stars. IR variability in debris disk if ongoing mining. |
| Populations (2.3) | Large numbers, masses | Small fraction of Main Belt mass. Feasible for stellar system wide economy in ~1000 yr interval (1% of journey time.) Ongoing and expanded surveys (e.g. VRO/LSST) for more interstellar asteroids will determine population density. |
| Controlled (3) | | |
| Where did it come from? (3.1) | Nearby stars. (3.1.1) | Large course correction needed. Calculate delta-v needed for each plausible star. |
| | T, L stars (3.1.2) | UKIDSS/WISE data appear to exclude. |

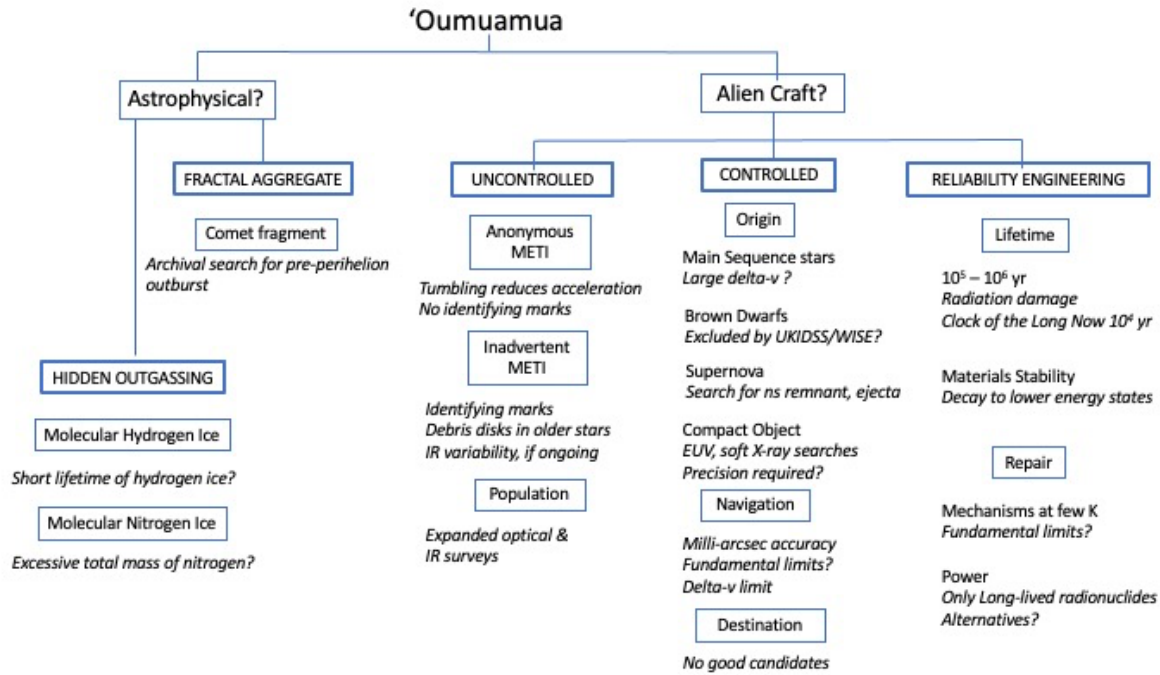

Figure 1: Astrophysical and Alien Craft explanations for 'Oumuamua and similar objects

**REFERENCES**

Adams, F.C., 2010, The Birth Environment of the Solar System, *Ann. Rev. Astronomy &Astrophysics*, 48, 47.

Bayler-Jones, C.A.L., et al., 2018, Plausible home stars of the interstellar object `Oumuamua found in Gaia DR2, *Astrophysical Journal*, 156, 205.

Bernstein, R.A., Freedman, W.L., & Madore, B.F., 2002, The First Detections of the Extragalactic Background Light at 3000, 5500, and 8000 Å. I. Results, *Astrophysical Journal*, 571, 56.

Best, W.M.J., et al., 2018, Photometry and Proper Motions of M, L, and T Dwarfs from the Pan-STARRS1 $3\pi$ Survey, *Astrophysical Journal Supplement*, 234, 1.

Bialy S., & Loeb, A., 2018, Could Solar Radiation Pressure Explain 'Oumuamua's Peculiar Acceleration? *Astrophysical Journal Letters*, 868, L1.

Bignami, G.F., 1996, Isolated Neutron Stars, *Science*, 271, 1372.

Butler, J.E., & Sumant, A.V., The CVD of Nanodiamond Materials, *Chemical Vapor Deposition*, 2008, 14, 145.

Do, A., Tucker, M.A., and Tonry, J., 2018, Interstellar Interlopers: Number Density and Origin of 'Oumuamua-like Objects, *Astrophysical Journal* Letters, 855, 10.

Dodd, M.S., et al., 2017, Evidence for early life in Earth's oldest hydrothermal vent precipitates, *Nature*, 543, Issue 7643, 60-64.

Drahus, M., et al., 2018, Tumbling motion of 1I/'Oumuamua and its implications for the body's distant past, *Nature Astronomy*, 2, 407.

Dressing, C.D. & Charbonneau, D., 2015, The Occurrence of Potentially Habitable Planets Orbiting M Dwarfs Estimated from The Full Kepler Dataset and an Empirical Measurement of the Detection Sensitivity, *Astrophysical Journal*, 807, 45.

Dybczyński, P.A., & Królikowska, M., 2018, Investigating the dynamical history of the interstellar

object 'Oumuamua, *Astronomy & Astrophysics*, 610, L11.

Ehresmann, M., 2021, Asteroid control through surface restructuring, *Acta Astronautica*, 178, 672.

Elvis, M., and Milligan, T., 2019, How much of the Solar System should we leave as Wilderness? *Acta Astronautica*, 162, 574.

Faucher-Giguère, C.-A., & Kaspi, V.M., 2006, Birth and Evolution of Isolated Radio Pulsars, *Astrophysical Journal*, 643, 332.

Fender, R.P., Maccarone, T.J., and Heywood, I., 2013, The closest black holes, *Monthly Notices of the Royal Astronomical Society*, 430, 1538.

Fesen, R.A., Weil, K.E., Cisneros, I.A., Blair, W.P., and Raymond, J.C., 2018, The Cygnus Loop's distance, properties, and environment driven Morphology, *Monthly Notices of the Royal Astronomical Society*, 481, 1786.

Fesen, R.A., et al., 2021, Far UV and Optical Emissions from Three Apparent Supernova Remnants Located at Unusually High Galactic Latitudes, arXiv:2102.12599.

Forgan, D.H., and Elvis, M., Extrasolar Asteroid Mining as Forensic Evidence for Extraterrestrial Intelligence, *International Journal of Astrobiology*, 10, 307.

Fraser, W.C., et al., 2018, The tumbling rotational state of 1I/'Oumuamua, *Nature Astronomy*, 2, 383.

Hallatt, T., & Wiegert, P., 2020, The Dynamics of Interstellar Asteroids and Comets within the Galaxy: an Assessment of Local Candidate Source Regions for 1I/'Oumuamua and 2I/Borisov, *Astrophysical Journal*, 159,147.

Hibberd, A., and Hein, A.M., 2021, Project Lyra: Catching 1I/'Oumuamua–Using Nuclear Thermal Rockets, *Acta Astronautica*, 179,594.

Hoang, T. and Loeb, A., 2020, Can Planet Nine be Detected Gravitationally by a Sub-Relativistic Spacecraft?, *Astrophysical Journal* Letters, 899, 23.

Jackson, A.P., and Desch, S. J., 2021, To see a World in a Shard of Ice: 'Oumuamua As A Fragment Of N2 Ice From An Exo-Pluto, *52nd Lunar and Planetary Science Conference 2021*, 1718.pdf

Johnson, R.C., 2003, The Slingshot Effect, https://maths.dur.ac.uk/%7edma0rcj/Psling/sling.pdf (accessed 27 April 2021.)

Katz, J.I., 2019, Evidence against non-gravitational acceleration of 1I/2017 U1 'Oumuamua, *Astrophys. Space Sci.*, 364, 51.

Katz, J.I., 2021, 'Oumuamua is not Artificial, arXiv:2102:07871.

Kessler D.J., and Cour-Palais, B.G., 1978, Collision Frequency of Artificial Satellites: The Creation of a Debris Belt, *Journal of Geophysical Research*, 83, 2637.

Lakatos, I., 1980, "The Methodology of Scientific Research Programmes", *Cambridge University Press*. ISBN-13: 978-0521280310.

Levine, W.G., Cabot, S.H.C., Seligman, D., and Laughlin, G., 2021, Constraints on the Occurrence of 'Oumuamua-Like Objects, ApJ, in press. arXiv:2108.11194.

Loeb, A., 2018, Six Strange Facts about our First Interstellar Guest, `Oumuamua, *Scientific American*, November 20. https://blogs.scientificamerican.com/observations/6-strange-facts-about-the-interstellar-visitor-oumuamua/ [accessed 27 April 2021.]

Loeb, A., "Extraterrestrial: The First Sign of Intelligent Life Beyond Earth", 2021, *Houghton Mifflin Harcourt*, New York, New York.

Lingam, M., and Loeb, A., 2019, Electric sails are potentially more effective than light sails near most stars, Acta Astronautica, 168, 146.

Luu, J.X., Flekkøy, E.G., and Toussaint, R., 2020, 'Oumuamua as a Cometary Fractal Aggregate: The "Dust Bunny" Model, *Astrophysical Journal* Letters, 900, L22.

Mancini Pires, A., 2017, What will eROSITA reveal among X-ray faint isolated neutron stars? *IAUS 337: Pulsar Astrophysics - The Next 50 Years*, eds: P. Weltevrede, B.B.P. Perera, L. Levin Preston & S. Sanidas. arXiv:1711.05038.

Mamajek, E.E., 2015, A Pre-Gaia Census of Nearby Stellar Groups, in *Young Stars & Planets Near the Sun*, Proceedings of IAU Symposium No. 314, J. H. Kastner, B. Stelzer, & S. A. Metchev, eds., p.21.

Mamajek, E.E., 2017, Kinematics of the Interstellar Vagabond 1I/`Oumuamua (A/2017 U1), *AAS Research Notes*, Nov 23.

Maoz, D., Ofek, E.O., and Shemi, A., 1997, Evidence for a new class of extreme ultraviolet sources, *Monthly Notices of the Royal Astronomical Society*, 287, 293.

Mashchenko, S., 2019, Modeling the light curve of 'Oumuamua: evidence for torque and disc-like shape, MNRAS, 489, 3003.

Matrá, L., 2020, Dust Populations in the Iconic Vega Planetary System Resolved by ALMA, *Astrophysical Journal*, 898, 146.

Matsuoka, Y., Ienaka, N., Kawara, K., and Oyabu, S., 2011, Cosmic Optical Background: The View from Pioneer 10/11, *Astrophysical Journal*, 736, 119.

Meech K.J., et al. 2017, A brief visit from a red and extremely elongated interstellar asteroid, *Nature*, 552, 378.

Micheli, M., et al., 2018, Non-gravitational acceleration in the trajectory of 1I/2017 U1 ('Oumuamua), *Nature*, 559, 223.

Moro-Martín, A., Turner, E.L., and Loeb, A., 2009, Will the Large Synoptic Survey Telescope Detect Extra-Solar Planetesimals Entering the Solar System? *Astrophysical Journal*, 704, 733.

Moro-Martin, A., 2019, Could 1I/'Oumuamua be an Icy Fractal Aggregate? *Astrophysical Journal Letters*, 832, L32.

O'Connor, P., and Kleyner, A., "Practical Reliability Engineering, 5th Edition", 2012, *Wiley*.

Phan, V.H.M., Hoang, T., & Loeb, A., 2021, Erosion of Icy Interstellar Objects by Cosmic Rays and Implications for `Oumuamua, arXiv:2019.04494.

Parry, L.A., et al., 2017, Ichnological evidence for meiofaunal bilaterians from the terminal Ediacaran and earliest Cambrian of Brazil, *Nature Ecology & Evolution*, 1, 1455.

Predehl, P., 2017, eROSITA on SRG, *Astronomische Nachrichten*, 338, 159.

Rafikov, R.R., 2018, Spin Evolution and Cometary Interpretation of the Interstellar Minor Object 1I/2017 'Oumuamua, *Astrophysical Journal Letters*, 867, L17.

Sartore N, Ripamonti, E., Treves, A., and Turolla, R., 2011, Space and velocity distributions of neutron stars in the Milky Way, *Adv. Sp. Research*, 47, 1294.

Schopf, J. W., et al., 2017, An anaerobic ∼3400 Ma shallow-water microbial consortium: Presumptive evidence of Earth's Paleoarchean anoxic atmosphere, *Precambrian Research*, 299, 309.

Sekanina, Z., 2019, 1I/'Oumuamua as Debris of Dwarf Interstellar Comet that Disintegrated before Perihelion, arXiv:1901.08704.

Seligman, D., Laughlin, G., and Batygin, K., 2019, On the Anomalous Acceleration of 1I/2017 U1 'Oumuamua, *Astrophysical Journal* Letters, 876, 26.

Setton, R., Bernier, P., and Lefrant, S., 2002, "Carbon Molecules and Materials", CRC Press.

Siraj, A., and Loeb, A., 2021, The Mass Budget Necessary to Explain `Oumuamua as a Nitrogen Iceberg, arXiv:2103.14032

Sleep, N.H., Zahnle, K.J., and Lupu, R.E., 2014, Terrestrial aftermath of the Moon-forming impact, *Proc. Roy. Soc. A*, 372, 1.

Sperber, B., 2019, "Bulk and Surface Related Degradation Phenomena in Monocrystalline Silicon at Elevated Temperature and Illumination", PhD thesis, Universität Konstanz. URL: http://nbn-resolving.de/urn:nbn:de:bsz:352-2-af03ujroa2ux1.y[p

Trilling, D.E., et al., 2018, Spitzer Observations of Interstellar Object 1I/'Oumuamua, *Astronomical Journal*, 156, 261.

Van Eylen, V., Agentoft, C., Lundkvist, M.S., Kjeldsen, H., Owen, J. E., Fulton, B. J., Petigura E., and Snellen, I., 2018, An asteroseismic view of the radius valley: stripped cores, not born rocky, Monthly Notices of the Royal Astronomical Society, 479, 4786-4795.

van Leeuwen, F., 2007, Validation of the new Hipparcos reduction, *Astronomy & Astrophysics*, 474, 653.

Weissman, P.R., 1983, The Mass of the Oort Cloud, *Astronomy & Astrophysics*, 118, 90.

Woosley, S.E., Heger, A., and Weaver, T.A., 2002, The evolution and explosion of massive stars, *Rev. Mod. Phys.*, 74, 1015.

Yefremov, A.P., 2020, Sensitivity of the Gravity Assist to Variations of the Impact Parameter, *Gravitation and Cosmology*, 26, 118.

Yoon, J., Peterson, D.M., Kurucz, R.L., and Zagarello, R.J., 2010, A New View of Vega's Composition, Mass, and Age, *Astrophysical Journal*, 708, 71.

Zuluaga, J.I., et al., 2018, A General Method for Assessing the Origin of Interstellar Small Bodies: The Case of 1I/2017 U1 ('Oumuamua)*, Astronomical Journal*, 155, 236.